\documentclass{Interspeech}

\interspeechcameraready

\title{ASDA: Audio Spectrogram Differential Attention Mechanism for Self-Supervised Representation Learning}

\author[affiliation={1}]{Junyu}{Wang}
\author[affiliation={1}]{Tianrui}{Wang}
\author[affiliation={1}]{Meng}{Ge}
\author[affiliation={1,2}]{Longbiao}{Wang}
\author[affiliation={3}]{Jianwu}{Dang}

\affiliation{Laboratory of Cognitive Computing and Application, College of Intelligence and Computing}{Tianjin University}{China}
\affiliation{Huiyan Technology (Tianjin) Co., Ltd}{Tianjin}{China}
\affiliation{Shenzhen Institute of Advanced Technology}{Chinese Academy of Sciences}{China}
\email{junyu\_wang21@tju.edu.cn}

\usepackage{comment}

\usepackage{amsmath}  
\usepackage{bm}       

\usepackage[table]{xcolor} 
\usepackage{colortbl}      

\begin{document}

\maketitle

\begin{abstract}

In recent advancements in audio self-supervised representation learning, the standard Transformer architecture has emerged as the predominant approach, yet its attention mechanism often allocates a portion of attention weights to irrelevant information, potentially impairing the model’s discriminative ability. To address this, we introduce a differential attention mechanism, which effectively mitigates ineffective attention allocation through the integration of dual-softmax operations and appropriately tuned differential coefficients. Experimental results demonstrate that our ASDA model achieves state-of-the-art (SOTA) performance across multiple benchmarks, including audio classification (49.0$\%$ mAP on AS-2M, 41.5$\%$ mAP on AS20K), keyword spotting (98.3$\%$ accuracy on SPC-2), and environmental sound classification (96.1$\%$ accuracy on ESC-50). These results highlight ASDA's effectiveness in audio tasks, paving the way for broader applications.
    
\end{abstract}

\section{Introduction}

In recent years, self-supervised learning (SSL) has demonstrated remarkable potential across various domains, including computer vision, natural language processing, and audio signal processing, by leveraging pre-training tasks such as contrastive learning and masked prediction to extract supervisory signals inherent in the data itself \cite{MAE, devlin-etal-2019-bert, speechSSL}. Particularly in the representation learning of sequential audio data, such as speech and music, SSL methods have proven effective in mitigating the scarcity of labeled data, thereby introducing a novel paradigm for audio understanding tasks \cite{hubert, wav2vec2.0}.

In the field of audio SSL, early work \cite{SSAST} first demonstrated the effectiveness of pure Transformer architectures through masked reconstruction tasks. However, the quadratic complexity of self-attention poses a significant computational burden. To alleviate this problem, \cite{audiomae} proposed an efficient encoding strategy that processes only a small amount of unmasked tokens during encoding phase, significantly improving computational efficiency. Nevertheless, this approach still requires substantial parameters to model complex dependencies during the decoding phase. Inspired by data2vec2.0 \cite{data2vec2.0}, EAT \cite{EAT} introduced an asymmetric encoder-decoder framework that employs multiple lightweight convolutional layers as the encoder, effectively reconstructing contextualized target representations while maintaining computational efficiency.

At the core of the Transformer architecture \cite{transformer} lies the dot-product attention mechanism, which enables the capture of global dependencies across tokens within an input sequence. Due to its advantages in parallel computation efficiency, long-range relationship modeling, and scalability, Transformer has rapidly evolved into a dominant neural network architecture in multiple domains. In audio processing, state-of-the-art (SOTA) SSL models \cite{audiomae, EAT, beats} predominantly adopt Vision Transformer (ViT) \cite{vit} as the backbone network to learn generalizable audio representations. Nonetheless, recent studies \cite{transformerlimitation, transformerlimitation2} have revealed a fundamental limitation of the standard Transformer: its attention allocation mechanism frequently distributes a portion of attention weights to irrelevant contextual information, which we refer to as the noise portion, thereby impairing the model’s ability to capture critical features.

To address these challenges, this study introduces a differential attention mechanism designed to mitigate the intrinsic noise introduced by single softmax operations \cite{differentialtransformer}. Drawing inspiration from methodologies in the enhancement domain \cite{differentialenhancement1, noiseheadphone}, this mechanism employs a dual-softmax operation to suppress irrelevant information in a differential manner, thereby refining attention allocation and enhancing the model’s ability to extract meaningful contextual cues. Building upon this foundation, we propose the Audio Spectrogram Differential Attention (ASDA) model, whose key components include differential attention modules, a MAE framework, and a teacher-student model architecture. The student model updates its parameters based on the teacher model’s output, while the teacher model employs an exponential moving average (EMA) update strategy \cite{EMA}, analogous to the data2vec framework \cite{data2vec}.

During pre-training, to alleviate the problem of unstable learned features caused by the fact that the input features of the student model are only 20\% of the complete features seen by the teacher model, and to share the high computational burden of the teacher model for processing the complete inputs, we introduce a multi-student single-teacher architecture. This design deploys multiple student models with distinct masked input positions under a shared teacher model, leveraging the relatively lower computational overhead of student models to achieve performance gains with minimal additional cost. Experimental results on multiple widely used audio benchmark datasets demonstrate that the proposed ASDA model consistently outperforms existing audio SSL approaches, achieving SOTA performance.

\begin{figure*}[t]
  \centering
  \vspace{-0.2cm}
  \includegraphics[width=0.7\linewidth]{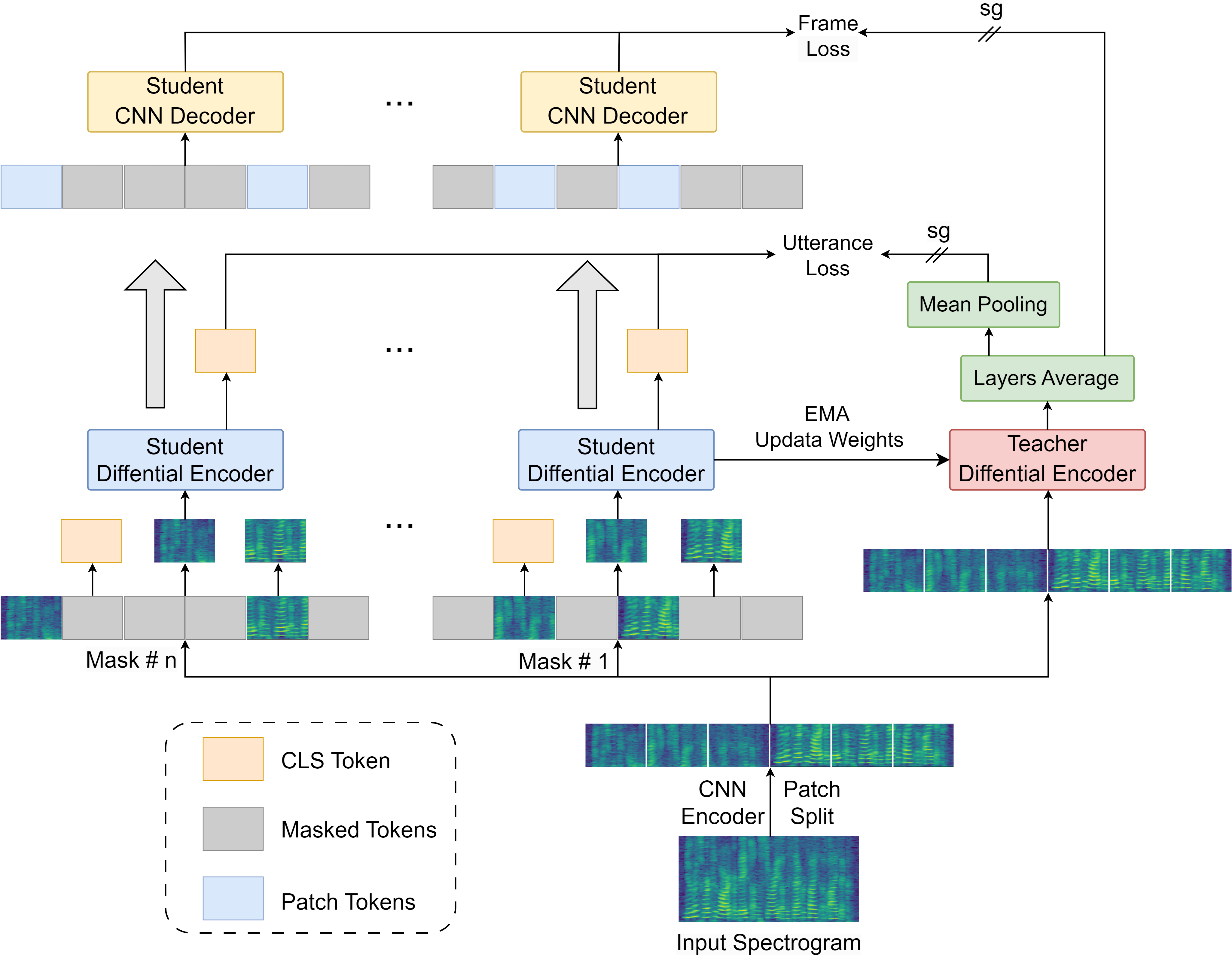}
  \caption{The overall architecture of the proposed ASDA model for self-supervised learning.}
  \label{figure1}
  \vspace{-0.2cm}
\end{figure*}

\section{Method}

\subsection{Model architecture}

The overall architecture of the proposed ASDA model is shown in Figure \ref{figure1}. Given a raw audio signal of approximately t seconds, we first convert it into a 128-dimensional log-mel filterbank (fbank) representation. Specifically, a 25 ms Hamming window is applied every 10 ms, yielding an input spectrogram of shape $128 \times 100t$. This spectrogram is then passed through a 2D convolutional layer to obtain the initial feature embeddings, followed by a patching operation that segments the embeddings into non-overlapping $16 \times 16$ patches. Each patch is subsequently flattened and projected into a 768-dimensional vector via a linear transformation, forming the patch embeddings \(X \in \mathbb{R}^{50t \times 768}\).

Since the Transformer architecture lacks an inherent mechanism for positional encoding, we incorporate fixed one-dimensional positional encodings into these embeddings to provide essential spatial awareness of the two-dimensional spectrogram representation. The patch embeddings are then fed into both the student and teacher models. For the student model, we employ a block-wise random masking strategy, as described in \cite{data2vec2.0}, and only the unmasked patches are used as input. To enhance the extraction of utterance-level information, we replace the average pooling operation with a learnable classification token (CLS token) similar to ViT \cite{vit}. These patch embeddings are processed by the student differential encoder, after which the masked segments are reintroduced, forming the complete representation that serves as input to the student CNN decoder, which predicts the frame-level spectrogram reconstruction. 

In contrast, the teacher model receives the full (unmasked) patch embeddings as input. These embeddings pass through the teacher differential encoder, producing differential attention outputs at each layer. Notably, the teacher differential encoder shares an identical architectural design with its student counterpart, ensuring consistent feature representations across both models and making them better for parameter adjustments via the EMA strategy.

\subsection{Differential attention}

The differential attention mechanism is inspired by the working principles of noise-canceling headphones \cite{differentialenhancement1, noiseheadphone}, where the core idea is to enhance informative acoustic signals while suppressing irrelevant noise through the optimized configuration of a parameter $\lambda$. Specifically, for the single-head attention mechanism, given an input feature matrix \(Z \in \mathbb{R}^{L \times D}\), we first apply linear transformations to obtain the query, key, and value representations. To achieve effective suppression of extraneous noise, we introduce a dual-path query-key mapping mechanism, mathematically formulated as follows:
\begin{figure}[t]
  \centering
  \vspace{-0.2cm}
  \includegraphics[width=0.63\linewidth]{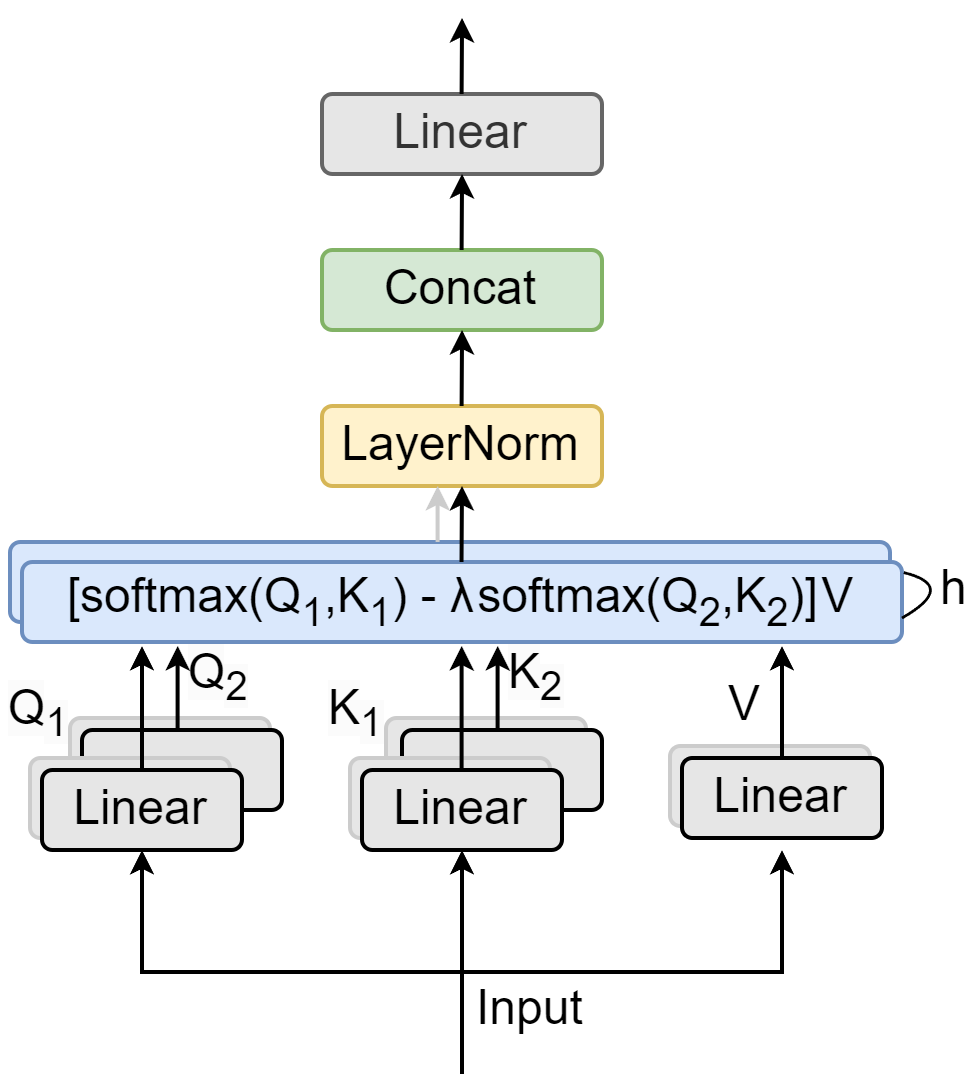}
  \caption{Differential attention Module.}
  \label{figure2}
  \vspace{-0.3cm}
\end{figure}
\begin{equation}
\begin{aligned}
[Q_1, Q_2] = ZW_Q, \thickspace [K_1, K_2] = ZW_K, \thickspace V = ZW_V \\
\end{aligned}
\end{equation}
where \(W_Q \in \mathbb{R}^{D \times 2D'}\), \(W_K \in \mathbb{R}^{D \times 2D'}\), and \(W_V \in \mathbb{R}^{D \times D'}\) are learnable parameter matrices. The differential attention weights are computed as:
\begin{equation}
\begin{aligned}
\textnormal{Diff}(Z) = \text{softmax}\left(\frac{Q_1K_1^{T}}{\sqrt{d}}\right) - \lambda &\hspace{0.2em}\text{softmax}\left(\frac{Q_2K_2^{T}}{\sqrt{d}}\right)  \\
\end{aligned}
\end{equation}
where $d$ denotes the feature dimension, and $\lambda$ is a tunable differential coefficient that controls the strength of noise suppression.

For the multi-head attention mechanism, differential attention is computed independently for each attention head. The outputs are then fused through layer normalization and concatenation, enabling multi-scale feature integration:
\begin{equation}
\begin{aligned}
\textnormal{head}_i = \textnormal{LayerNorm}(\textnormal{Diff}_i(Z)V), \hspace{1em} i& \in [1,h] \\
\textnormal{MultiHead}(Z) = \textnormal{Concat}(\textnormal{head}_1, \textnormal{head}_2, .&.., \textnormal{head}_h)W_O\\
\end{aligned}
\end{equation}
where \(W_O \in \mathbb{R}^{D' \times D}\) is the output projection matrix and $h$ denotes the number of attention heads, which is set to 8 in our experiments. Figure \ref{figure2} illustrates the overall structure of the differential attention.

\subsection{Pre-training and fine-tuning details}

The model comprises 95M and 93M trainable parameters during the pre-training and fine-tuning stages, respectively. During pre-training, both the student and teacher differential encoders adopt a 12-layer architecture, each consisting of stacked differential attention modules and feed-forward networks (FFNs). Each FFN consists of two fully connected layers with GeLU activation functions \cite{gelu}. The student CNN decoder is composed of six layers of 2D convolutions, followed by LayerNorm and GeLU activation functions.

To enhance the global modeling capability of the encoder, we introduce an additional contrastive loss term before passing the encoder output to the decoder. Unlike conventional self-supervised audio learning methods that rely solely on frame-level reconstruction loss, we compute a contrastive loss between the CLS token representation from the student model and the global average pooled representation from the teacher model, following a strategy similar to \cite{EAT}.

During the fine-tuning stage, only the student encoder is retained, while the teacher model and CNN decoder are removed. The input masking ratio is set to 0.2, balancing regularization and computational efficiency, a configuration empirically validated as effective in our experiments. Finally, a trainable linear classification layer is added on top of the encoder, mapping the learned abstract representations to the target label space, enabling efficient transfer learning for downstream tasks.

\subsection{Loss function}

The utterance-level loss quantifies the discrepancy between the CLS representation from the student encoder and the global representation derived from the multi-layer teacher encoder outputs:
\begin{equation}
\begin{aligned}
 \mathcal{L}_{\text{utterance}} = \vert \vert \mathbf{Y}'_s - \text{GAP}\left(\frac{1}{L}\sum_{l=1}^{L} \mathbf{Y}'_{t,l}\right) \vert\vert_2^2
\end{aligned}
\end{equation}
where \(Y'_s \in \mathbb{R}^{1 \times D}\) represents the CLS token output from the student encoder, \(Y'_{t,l} \in \mathbb{R}^{T \times D}\) denotes the feature representation at the 
l-th layer of the teacher encoder, and GAP(·) refers to the global mean pooling operation. The total number of encoder layers is denoted as \(L\).

The frame-level loss measures the discrepancy between the spectrogram reconstructed by the CNN decoder and the original spectrogram output from the teacher encoder:
\begin{equation}
\begin{aligned}
\mathcal{L}_{\text{frame}} =\left \vert\vert \mathbf{Y}_s - \mathbf{Y}_t \right \vert\vert_2^2
\end{aligned}
\end{equation}
where \(Y_s \in \mathbb{R}^{T \times F}\) represents the predicted spectrogram from the CNN decoder, \(Y_{t} \in \mathbb{R}^{T \times F}\) corresponds to the target spectrogram produced by the teacher encoder.

The overall loss function is defined as the weighted sum of the utterance-level and frame-level losses:
\begin{equation}
\begin{aligned}
\mathcal{L}_{\text{total}} = \alpha \cdot \mathcal{L}_{\text{utterance}} + \mathcal{L}_{\text{frame}}
\end{aligned}
\end{equation}
where $\alpha$ is a tunable hyperparameter, which balances the learning of global utterance-level representations and local frame-level spectral details.

\section{Experiments} 

Our study leverages the large-scale AudioSet dataset \cite{audioset} for model pre-training and evaluates its performance across three representative downstream tasks: audio classification (AS-2M and AS20K), keyword spotting (Speech Commands V2) \cite{speechcommand}, and environmental sound classification (ESC-50) \cite{ESC-50}.

\subsection{Datasets}

The AudioSet dataset comprises approximately 2 million 10-second audio clips spanning 527 sound categories. To ensure a fair comparison with existing methods, we utilize 1,912,134 samples for pre-training and fine-tuning on AS-2M, while 20,550 samples are allocated for fine-tuning on AS20K. Given the multi-label nature of the dataset, we adopt mean Average Precision (mAP) as the primary evaluation metric.

For speech-related tasks, we employ the Speech Commands V2 (SPC-2) dataset, which consists of approximately 105K 1-second utterances across 35 commonly used speech commands. The dataset is pre-divided into training (84,843 samples), validation (9,981 samples), and test sets (11,005 samples), and we follow the official split for evaluation.

In environmental sound classification, we utilize the ESC-50 dataset, which contains 2,000 5-second audio clips distributed across 50 environmental sound categories. Due to the relatively small dataset size, we employ a five-fold cross-validation strategy to obtain a more robust and reliable assessment of model performance.

\begin{table*}[ht]
\begin{center}
  \caption{Performance comparison with existing methods across multiple audio tasks. The symbol “-” indicates that the data was not reported in the original paper. “Acc” represents accuracy, which is used as the evaluation metric for single-label classification tasks. Regarding pre-training datasets, "AS" refers to AudioSet, "LS" denotes LibriSpeech, and "IN" corresponds to ImageNet. Additionally, methods that incorporate extra supervised training or leverage auxiliary labeling tasks are highlighted in grey for clarity.}
  \setlength{\tabcolsep}{2mm}
  \label{tab1}
  \centering
  \vspace{-0.1cm}
  \begin{tabular}{lcccccc}
    \hline
    \textbf{Model}  & \textbf{Data} & \textbf{\# Params}  & \textbf{AS-2M (mAP)}  & \textbf{AS20K (mAP)}  & \textbf{SPC-2 (Acc.)}   & \textbf{ESC-50 (Acc.)}   \\
    \hline
    \textbf{Supervised pre-training} \\
    \textcolor{gray}{AST \cite{ast}} & \textcolor{gray}{IN} & \textcolor{gray}{86M} & \textcolor{gray}{45.9} & \textcolor{gray}{34.7} & \textcolor{gray}{88.7} & \textcolor{gray}{98.1}              \\
    \textcolor{gray}{MBT \cite{mbt}} & \textcolor{gray}{IN-21K} & \textcolor{gray}{86M} & \textcolor{gray}{44.3} & \textcolor{gray}{31.3} & \textcolor{gray}{-} & \textcolor{gray}{-}              \\
    \textcolor{gray}{PaSST \cite{passt}} & \textcolor{gray}{IN} & \textcolor{gray}{86M} & \textcolor{gray}{47.1} & \textcolor{gray}{-} & \textcolor{gray}{-} & \textcolor{gray}{96.8}              \\
    \hline
    \textbf{Self-supervised pre-training} \\
    Conformer \cite{conformerSSL} & AS & 88M & 41.1 & - & - & 88.0      \\
    SS-AST \cite{SSAST} & AS+LS & 89M & - & 31.0 & 98.0 & 88.8      \\
    data2vec \cite{data2vec} & AS & 94M & - & 34.5 & - & -      \\
    Audio-MAE \cite{audiomae} & AS & 86M & 47.3 & 37.1 & \textbf{98.3} & 94.1      \\
    BEATs \cite{beats} & AS & 90M & 48.0 & 38.3 & \textbf{98.3} & 95.6      \\
    EAT \cite{EAT} & AS & 88M & 48.6 & 40.2 & \textbf{98.3} & 95.9      \\
    \hline
    ASDA & AS & 93M & \textbf{49.0} & \textbf{41.5} & \textbf{98.3} & \textbf{96.1}      \\
    \hline
  \end{tabular}
  \vspace{-0.5cm}
\end{center}
\end{table*}

\subsection{Experimental setup}

The proposed model architecture incorporates 16 student networks (\(n=16\)), with a CNN-based student decoder designed using grouped convolutions, consisting of 16 groups of 3×3 2D convolutional filters. Model training is conducted on four NVIDIA 4090 GPUs using a distributed data parallel strategy. During pre-training, the model is trained for 20 epochs with a batch size of 48. We adopt the Adam optimizer \cite{adam}, with hyperparameters set as $\beta_1$ = 0.9, $\beta_2$ = 0.95, and a weight decay coefficient of 0.05. The learning rate is scheduled using a cosine annealing strategy with warm-up, where the peak learning rate is set to 5e-4, and the warm-up phase spans approximately 2.5 epochs to ensure training stability in the early stages.

\section{Results}

\subsection{Performance comparison on standard benchmarks}

Table \ref{tab1} presents the performance comparison between our model and various classical baseline methods. The experimental results demonstrate that compared to the current best-performing extra-supervised pre-training model \cite{passt}, our method achieves a significant improvement of 1.9$\%$ mAP on the large-scale audio dataset AS-2M, while only slightly underperforming by 0.7$\%$ accuracy on the small-scale environmental sound classification dataset ESC-50. To ensure a fair comparison, we primarily focus on comparing with other self-supervised pre-training methods.

Specifically, in audio classification tasks, our method achieves improvements of 0.4$\%$ and 1.3$\%$ mAP on AS-2M and AS20K datasets, respectively, significantly surpassing previous SOTA models EAT and BEATs. In environmental sound classification, our method reaches an accuracy of 96.1$\%$ on the ESC-50 dataset, setting a new SOTA performance.

Furthermore, although our experiments mainly focus on audio tasks, we also validate our model on speech-related tasks. On the keyword spotting dataset SPC-2, our method achieves an accuracy of 98.3$\%$, which is the same as previous SOTA results. These results indicate that our ASDA model exhibits excellent generalization capability in modeling both audio and speech tasks.

\begin{table}[h]
  \vspace{-0.3cm}
  \caption{Performance comparison of loss weight $\alpha$ and the impact of CLS token placement and pooling strategy.}
  \label{tab2}
  \centering
  \vspace{-0.1cm}
  \tabcolsep=0.11cm  
  \begin{tabular*}{\hsize}{@{\hspace{0.15cm}}@{\extracolsep{\fill}}cccc@{\hspace{0.3cm}}}
    \hline
    \textbf{Method}  & \textbf{AS20K} & \textbf{SPC-2} & \textbf{ESC-50} \\
    \hline
    \textbf{Loss weight} \\
    $\alpha$=0.5  &  \textbf{41.5} & \textbf{98.3} & \textbf{96.1} \\
    $\alpha$=1 & 41.3 & \textbf{98.3} & 96.0 \\
    $\alpha$=2 & 41.1 & \textbf{98.3}  & 96.0 \\
    \hline
    \textbf{Classfication strategy} \\
    Head CLS token  &  \textbf{41.5} & \textbf{98.3} & \textbf{96.1} \\
    Middle CLS token & 41.1 & \textbf{98.3} & 95.9 \\
    Mean pooling & 41.1 & \textbf{98.3} & 96.0 \\
    \hline
  \end{tabular*}
  \vspace{-0.2cm}
\end{table}

\subsection{Ablation studies}

Table \ref{tab2} presents the impact of different loss weights and classification strategies on model performance. The experimental results demonstrate that when the hyperparameter $\alpha$ is set to 0.5, the model achieves an optimal balance between utterance-level and frame-level feature learning capabilities. Furthermore, by incorporating the utterance loss, the feature extraction ability of the CLS token is further improved compared to the traditional mean pooling approach.

We also investigate the influence of different positional placements of the CLS token on model performance. The experimental data reveal that the head CLS token outperforms the middle CLS token, which aggregates information bidirectionally from both the beginning and end of the sequence. We hypothesize that this phenomenon may be attributed to the following reason: when the CLS token is placed in the middle of the sequence, the distribution of its attention weights is subject to bidirectional interference from preceding and subsequent tokens, leading to increased instability in the information aggregation process and consequently degrading the quality of the final representation. In contrast, the unidirectional information aggregation mechanism of the head CLS token enables more stable aggregation of global sequence information.

\begin{table}[h]
  \vspace{-0.3cm}
  \caption{The effect of different differential coefficients $\lambda$ on model performance in AS20K.}
  \label{tab3}
  \centering
  \vspace{-0.1cm}
  \tabcolsep=0.11cm  
  \begin{tabular*}{\hsize}{@{\hspace{0.15cm}}@{\extracolsep{\fill}}ccccc@{\hspace{0.3cm}}}
    \hline
    \textbf{Model} & $\lambda$=0 & $\lambda$=0.1 & $\lambda$=0.3 & $\lambda$=0.5  \\
    \hline
    \textbf{AS20K} & 41.0 & 41.4 & \textbf{41.5} & 41.1 \\
    \hline
  \end{tabular*}
  \vspace{-0.2cm}
\end{table}

Table \ref{tab3} presents an analysis of the impact of different differential coefficients $\lambda$ on model performance. Here, $\lambda$=0 indicates the absence of the differential attention mechanism, where the model structure resembles the standard ViT architecture. The experimental results demonstrate that the differential attention mechanism significantly enhances model performance. By appropriately setting the value of the differential coefficient $\lambda$, noticeable performance improvements can be achieved without altering the overall model architecture. This finding strongly aligns with our vision of providing a universal foundational architecture for SSL in audio processing, validating the effectiveness and practicality of the differential attention mechanism in this domain.

\section{Conclusions}
In this paper, we introduce a novel differential attention mechanism to address the issue of standard Transformer architectures allocating excessive attention weights to irrelevant contextual information. By defining such irrelevant information as noise and drawing inspiration from differential denoising techniques, we design a dual-softmax based differential attention mechanism. This mechanism effectively eliminates noise interference while preserving useful information through appropriate differential operations. Building upon this, we integrate a teacher-student framework to further enhance the model’s capability in extracting critical features. Experimental results demonstrate that the proposed ASDA model establishes new state-of-the-art (SOTA) performance across multiple benchmark datasets in audio and speech processing. In future work, we aim to extend the differential attention mechanism to more challenging audio-speech joint training scenarios, further exploring its potential in multimodal learning and providing a generalizable foundational framework for a broader range of audio processing tasks.

\bibliographystyle{IEEEtran}
\bibliography{mybib}

\end{document}